\documentclass[aip,jcp,amsmath,preprint,reprint]{revtex4-1}
\usepackage{amsmath}
\usepackage{dcolumn}

\renewcommand{\vec}[1]{{\rm\bf #1}}
\newcommand{\frct}[2]{{\textstyle\frac{#1}{#2}}}

\newcommand{\ep}{\epsilon}
\newcommand{\de}{{\rm d}}
\newcommand{\im}{{\rm i}}
\newcommand{\sm}{\boldsymbol\sigma}
\newcommand{\hh}{\hat{\mathcal{H}}}

\begin{document}

\title{Additive atomic approximation for relativistic effects:
a two-component Hamiltonian
for molecular electronic structure calculations}

\author{Dimitri N. Laikov}
\email[E-Mail: ]{laikov@rad.chem.msu.ru}
\homepage[Homepage: ]{http://rad.chem.msu.ru/~laikov/}
\affiliation{Chemistry Department, Moscow State University,
119991 Moscow, Russia}

\date{\today}

\begin{abstract}
An approximate relativistic two-component Hamiltonian
for use in molecular electronic structure calculations
is derived in the form of a sum
of fixed atom-centered kinetic and spin-orbit operators
added to the non-relativistic Hamiltonian.
Starting from the well-known zeroth-order regular approximation,
further steps are taken to get rid of its nonlinearity
in the potential, ending up with a simple formulation
with easily computable integrals that can seamlessly work
with any traditional electronic structure method.
Molecular tests show a good accuracy of this approximation.
\end{abstract}

\maketitle


Even though the Dirac~\cite{D28} equation
is the best theory of the electron,
its simple nonrelativistic approximation ---
the Schr\"odinger~\cite{S26} equation ---
is most widely used
in molecular electronic structure calculations.
When it comes to heavier atoms,
the effective core potentials~\cite{H35,KBT76}
most often come into play, as they can model,
albeit in a rather queer way,
the relativistic effects felt by the valence electrons;
besides their quite arbitrary parametrization,
they may suffer from numerical instability~\cite{W06}
and need a special care~\cite{HCD17}
in the evaluation of their integrals
over the traditional Gaussian~\cite{B50} functions.

Another two-component formalism,
until now limited only to density-functional~\cite{KS65} methods,
is the zeroth-order regular approximation~\cite{CPD86,LBS93,LBS94,S06}
that works with a local effective one-electron potential
in such a nonlinear way that it cannot deal
with the more general many-electron wavefunction theories.
We have found, however, a further approximation
that removes this nonlinearity
and leads to a very simple one-electron Hamiltonian
for molecular electronic structure theories,
the straightforward analytical evaluation of its matrix elements
over the Gaussian basis sets allows it to be easily implemented
into quantum chemistry codes for both correlated-wavefunction
and density-functional calculations.
We are glad to report it here,
along with the construction
of atomic basis sets of our~\cite{L05,L19}
correlation-consistent type.
We test its performance
on some characteristic molecular examples
using the coupled-cluster~\cite{C58,CK60,C66} theory
with single and double~\cite{PB82} (CCSD) substitutions.


For an electron in a potential $v(\vec{r})$
the Dirac equation can be written as
\begin{equation}
\label{eq:4c}
\left\{
\begin{array}{rcrcl}
\bigl( v(\vec{r}) - \ep \bigr) \psi_\textrm{L} (\vec{r}) &+& c\, \sm\cdot\vec{p}\, \psi_\textrm{S} (\vec{r}) &=& 0 \\
c\, \sm\cdot\vec{p}\, \psi_\textrm{L} (\vec{r}) &+& \bigl( v(\vec{r}) - 2c^2 - \ep \bigr) \psi_\textrm{S} (\vec{r})  &=& 0
\end{array}
\right.
\end{equation}
with the Pauli matrices
\begin{equation}
\sm = \left\{
\left(
\begin{array}{rr}
0 & 1 \\
1 & 0
\end{array}
\right),
\left(
\begin{array}{rr}
0   & -\im \\
\im & 0
\end{array}
\right),
\left(
\begin{array}{rr}
1 & 0 \\
0 & 1
\end{array}
\right)
\right\} .
\end{equation}
The relation between the small $\psi_\textrm{S} (\vec{r})$ and large $\psi_\textrm{L} (\vec{r})$
component wavefunctions from the second line of Eq.~(\ref{eq:4c})
\begin{equation}
\label{eq:ps}
\psi_\textrm{S} (\vec{r}) =
\frac{c}{2c^2 - v(\vec{r}) + \ep}
 \, \sm\cdot\vec{p}\, \psi_\textrm{L} (\vec{r})
\end{equation}
can be \textit{simplified} and \textit{approximated}
by dropping out the energy $\ep$ in the denominator of Eq.~(\ref{eq:ps}),
as for the valence electrons in molecules $\ep \ll 2c^2$
and a typically attractive effective potential $v(\vec{r}) < 0$.
Putting the approximate $\psi_\textrm{S} (\vec{r})$
into the first line of Eq.~(\ref{eq:4c})
leads to the two-component relativistic equation
\begin{equation}
\label{eq:zo}
\left( v(\vec{r})
 + \sm\cdot\vec{p}\, \frac{c^2}{2c^2 - v(\vec{r})} \sm\cdot\vec{p}
\right) \psi (\vec{r})
 = \ep \psi (\vec{r})
\end{equation}
that works only with the large component
$\psi (\vec{r}) \equiv \psi_\textrm{L} (\vec{r})$,
the well-known zeroth-order regular approximation~\cite{LBS93}.

The appearance of $v(\vec{r})$ in Eq.~(\ref{eq:zo})
in a non-linear way besides the usual multiplicative potential
limits the applications of this formalism
to the local-potential models of density-functional theory,
where the effective potential is solved for self-consistently
together with the one-electron wavefunctions it depends on.
Now we will make further approximations
to get rid of this non-linearity by making
the relativistic correction term independent
of the molecular electronic structure.

The Hamiltonian of Eq.~(\ref{eq:zo}) can be written
\begin{equation}
\label{eq:hu}
\hh = v(\vec{r})
 - \frct12 \nabla^2
 + \frct12 \sm\cdot\vec{p}\, u(\vec{r})\, \sm\cdot\vec{p}
= \hh_0 + \hh_u
\end{equation}
as a sum of the non-relativistic $\hh_0$
and the relativistic correction $\hh_u$
with the scalar function
\begin{equation}
\label{eq:ur}
u(\vec{r}) = \frac{v(\vec{r})}{2c^2 - v(\vec{r})}
\end{equation}
that for $v(\vec{r}) \le 0$
has the range $-1 \le u(\vec{r}) \le 0$
and plays a role mostly in the atomic core regions,
as $-v(\vec{r}) \ll 2c^2$ elsewhere.
We approximate $v(\vec{r})$ in Eq.~(\ref{eq:ur}) by a sum
of fixed non- (or weakly) overlapping spherically-symmetric
atom-centered potentials
\begin{equation}
\label{eq:vn}
\bar{v}(\vec{r}) =
\sum\nolimits_k v_k \bigl( |\vec{r} - \vec{r}_k| \bigr),
\end{equation}
then $u(\vec{r})$ can be given in the same way as
\begin{equation}
\label{eq:un}
u(\vec{r}) =
\sum\nolimits_k u_k \bigl( |\vec{r} - \vec{r}_k| \bigr),
\end{equation}
our localized atomic radial functions
\begin{equation}
\label{eq:uk}
u_k (r) = b(r) \frac{v_k(r)}{2c^2 - v_k(r)}
\end{equation}
further have a cut-off factor
\begin{equation}
\label{eq:b}
b(r) = \left\{
\begin{array}{lr}
1, & r < r_1 \\
\frct12 - \frct12 s\bigl((2r - r_0 - r_1)/(r_0 - r_1)\bigr), & r_1 \le r \le r_0 \\
0, & r > r_0
\end{array}
\right.
\end{equation}
that brings them smoothly to zero for $r > r_0$.

We take $v_k(r)$ in Eq.~(\ref{eq:uk})
as the purely electrostatic potential
of the nuclear charge and electron density
from the spherically-symmetric average-level~\cite{GMP76}
Hartree-Fock calculations on the neutral atoms
within the scalar-relativistic approximation~\cite{D94}
of the Dirac-Coulomb Hamiltonian --- as such,
it has an exponentially-decaying tail
that can be easily bent down to zero.
Had we added a local exchange potential,
there would have been either a very long tail
for an asymptotically correct $v_k(r)\to -1/r$ as $r\to\infty$,
or a mild deepening of $v_k(r)$ mostly where we are going
to cut it down.
Moreover, the lack of the exchange term may somewhat compensate
for the energy dropped in the denominator of Eq.~(\ref{eq:ps}).

In Eq.~(\ref{eq:b}) we use our favorite switching function
\begin{equation}
\label{eq:s}
s(x) = \tanh\left(\sqrt{3}\, \frac{x}{1-x^2}\right)
\end{equation}
where the factor of $\sqrt{3}$ makes the third derivative
$s'''(0) = 0$,
and we find the cut-off limits
$r_1 = \frct12$ and $r_0 = \frct32$
to yield the rather smooth $u_k(r)$ from Eq.~(\ref{eq:uk}), for all atoms,
without inflection points between $r_1$ and $r_0$.
We have tried to find a better cut-off factor,
for example, a more general function could be used
with parameters adjusted to minimize some criterion,
but it did not yield a better output of the whole work.

Our last step is to make the integrals
of the Hamiltonian of Eqs.~(\ref{eq:hu}) and~(\ref{eq:un})
over the traditional Gaussian basis sets
easy to compute analytically,
so we further approximate $u_k(r)$ from Eq.~(\ref{eq:uk})
by sums of Gaussians
\begin{equation}
\label{eq:ua}
u_k(r) \approx \tilde{u}_k(r) = \sum\nolimits_{i=1}^n c_{ik} \exp\left(-a_{ik} r^2 \right)
\end{equation}
with parameters $\{c_{ik}\}$ and $\{a_{ik}\}$
from the least-squares minimization
\begin{equation}
\label{eq:uu}
\int\limits_0^\infty
\bigl| \tilde{u}_k(r) - u_k(r) \bigr|^2 \de r = \min .
\end{equation}
We find $n = 15$ in Eq.~(\ref{eq:ua}) to be the best for all atoms,
it is the greatest $n$ for which all $c_{ik}$ are still negative,
the smallest $a_{1k}$ falls in the range $2.4 < a_{1k} < 3.6$
with $|c_{1k}| < 0.002$,
and the fit error
$\mu = \max\limits_r \bigl| \tilde{u}_k(r) - u_k(r) \bigr| < 0.0011$.
Another good $n = 17$ has $\mu < 0.00007$ thanks to a better fit
at a very short $r$ with one positive $c_{ik}$,
but it makes only a slight change in the computed atomic wavefunctions.
For comparison, we have also made
very accurate fits with $38 \le n \le 50$
for the uncut $u_k(r)$ with $b(r)\equiv1$ in Eq.~(\ref{eq:uk}),
and the atomic calculations show
that by far the greatest change comes
when the two-component Eq.~(\ref{eq:zo})
is used instead of the four-component Eq.~(\ref{eq:4c}),
the cut-off function of Eqs.~(\ref{eq:b}) and~(\ref{eq:s})
has then only a slight effect,
and even less so does the 15-term fit of Eq.~(\ref{eq:ua}).

The relativistic Hamiltonian term $\hh_u$ of Eq.~(\ref{eq:hu})
can be split into a ``scalar'' part
\begin{equation}
\label{eq:hs}
\hh_\textrm{s} =
 - \frct12 \nabla\cdot\, u(\vec{r})\, \nabla
\end{equation}
that can be thought of as a rescaling of the kinetic energy,
and a ``vector'' part
\begin{equation}
\hh_\textrm{v} =
 - \frct12 \im\sm\cdot \Bigl(\bigl(\nabla u(\vec{r})\bigr) \times\nabla\Bigr)
\end{equation}
that leads to the spin-orbit coupling.
With $\hh_\textrm{s}$ only, we get a scalar-relativistic approximation
that can seamlessly work with any traditional non-relativistic
electronic structure method to allow the studies of molecules
with heavier atoms, we would call it an additive atomic approximation (AAA),
and we implement it first into our computer code.
The spin-orbit term $\hh_\textrm{v}$ can be treated as a perturbation,
and we believe our function $u(\vec{r})$ to be a sound choice for this.

Throughout our work,
we use the newest estimate~\cite{AHKN12,HFG11} of the speed of light
$c=137.035999173$
and the finite nucleus model~\cite{VD97}
with Gaussian charge distribution with the exponent (in au)
\begin{equation}
\alpha = \frct32 \left(\frac{529177249}{5700 + 8360\cdot\sqrt[3]{M}}\right)^2
\end{equation}
where $M$ is the (integer) mass number of the most abundant isotope.


We run our atomic electronic structure code~\cite{L19}
working in 256-bit precision, first to get
the nearly-exact four-component scalar-relativistic~\cite{D94}
spherically-symmetric atomic Hartree-Fock solutions
over the huge even-tempered Gaussian basis sets,
for all 102 atoms from Hydrogen through Nobelium.
Then we compute the values of function $u_k(r)$ from Eq.~(\ref{eq:uk})
on a very dense grid of points
in the spirit of double-exponential integration~\cite{TM74},
and get the 15-term Gaussian fits of Eqs.~(\ref{eq:ua})
minimizing the integral of Eq.~(\ref{eq:uu})
computed numerically on the grid;
we check the goodness of the fit by plotting, for all atoms,
the difference $\tilde{u}_k(r) - u_k(r)$
as a function of $\ln(r)$ and also
the values of $\ln(a_{ik})$ against the atomic number,
and see that it is good.
These values of $\{c_{ik}\}$ and $\{a_{ik}\}$
are tabulated to 20 decimal places
in the supplementary material~\cite{SM}
and everyone is welcome to use them.

For our AAA Hamiltonian, we get our correlation-consistent
atomic basis sets, for all 102 atoms,
in the same way as we did it before~\cite{L19},
and we tabulate them in the supplementary material~\cite{SM}.
The molecular properties computed therewith
can be compared one-to-one with those based
on the four-component scalar-relativistic~\cite{D94} Hamiltonian,
and even more so for the longer underlying primitive sets.

\begingroup
\begin{table}
\caption{\label{tab:mol}Molecular properties computed by CCSD/L2\_4.}
\begin{ruledtabular}
\begin{tabular}{llrrcllrr}
mol. & $\hat{H}$\textsuperscript{a} & \multicolumn{1}{c}{$r$} & \multicolumn{1}{c}{$\Delta E$} &&
mol. & $\hat{H}$\textsuperscript{a} & \multicolumn{1}{c}{$r$} & \multicolumn{1}{c}{$\Delta E$} \\
\hline
LiH    & 4 & 3.0216 & 0.09070 && Li$_2$ & 4 & 5.0684 & 0.03785 \\
       & 2 & 3.0217 & 0.09070 &&        & 2 & 5.0685 & 0.03785 \\
       & n & 3.0218 & 0.09070 &&        & n & 5.0688 & 0.03785 \\[1pt]
NaH    & 4 & 3.5746 & 0.07000 && Na$_2$ & 4 & 5.8783 & 0.02564 \\
       & 2 & 3.5748 & 0.07000 &&        & 2 & 5.8789 & 0.02564 \\
       & n & 3.5775 & 0.07003 &&        & n & 5.8864 & 0.02563 \\[1pt]
 KH    & 4 & 4.2632 & 0.06406 &&  K$_2$ & 4 & 7.5392 & 0.01682 \\
       & 2 & 4.2635 & 0.06407 &&        & 2 & 7.5403 & 0.01682 \\
       & n & 4.2700 & 0.06426 &&        & n & 7.5664 & 0.01685 \\[1pt]
RbH    & 4 & 4.5019 & 0.06184 && Rb$_2$ & 4 & 8.1152 & 0.01418 \\
       & 2 & 4.5023 & 0.06185 &&        & 2 & 8.1169 & 0.01418 \\[1pt]
CsH    & 4 & 4.7620 & 0.06333 && Cs$_2$ & 4 & 9.0258 & 0.01201 \\
       & 2 & 4.7611 & 0.06335 &&        & 2 & 9.0263 & 0.01200 \\[2pt]
CuH    & 4 & 2.7694 & 0.09856 && Cu$_2$ & 4 & 4.2797 & 0.06390 \\
       & 2 & 2.7705 & 0.09846 &&        & 2 & 4.2813 & 0.06381 \\
       & n & 2.8195 & 0.09482 &&        & n & 4.3487 & 0.06023 \\[1pt]
AgH    & 4 & 3.0614 & 0.08442 && Ag$_2$ & 4 & 4.8571 & 0.05214 \\
       & 2 & 3.0636 & 0.08427 &&        & 2 & 4.8597 & 0.05200 \\[1pt]
AuH    & 4 & 2.8758 & 0.11323 && Au$_2$ & 4 & 4.7387 & 0.07089 \\
       & 2 & 2.8791 & 0.11220 &&        & 2 & 4.7461 & 0.06990 \\[2pt]
HF     & 4 & 1.7271 & 0.21599 &&  F$_2$ & 4 & 2.6481 & 0.03978 \\
       & 2 & 1.7270 & 0.21605 &&        & 2 & 2.6481 & 0.03979 \\
       & n & 1.7271 & 0.21631 &&        & n & 2.6476 & 0.03984 \\[1pt]
HCl    & 4 & 2.4098 & 0.16321 && Cl$_2$ & 4 & 3.7910 & 0.07483 \\
       & 2 & 2.4098 & 0.16324 &&        & 2 & 3.7910 & 0.07486 \\
       & n & 2.4100 & 0.16360 &&        & n & 3.7897 & 0.07516 \\[1pt]
HBr    & 4 & 2.6771 & 0.14198 && Br$_2$ & 4 & 4.3469 & 0.06580 \\
       & 2 & 2.6773 & 0.14199 &&        & 2 & 4.3473 & 0.06583 \\
       & n & 2.6814 & 0.14294 &&        & n & 4.3518 & 0.06666 \\[1pt]
HI     & 4 & 3.0513 & 0.12109 &&  I$_2$ & 4 & 5.0903 & 0.05866 \\
       & 2 & 3.0524 & 0.12105 &&        & 2 & 5.0921 & 0.05868 \\[2pt]
Sb$_2$ & 4 & 4.7403 & 0.07654 && UO$_3$ & 4 & 3.3565 &         \\
       & 2 & 4.7426 & 0.07650 &&        & 2 & 3.3462 &         \\
Bi$_2$ & 4 & 5.0248 & 0.05964 &&        & 4 & 3.4667 &         \\
       & 2 & 5.0394 & 0.05932 &&        & 2 & 3.4624 &         
\end{tabular}
\begin{flushleft}
\textsuperscript{a}Hamiltonian: 4~four-, 2~two-component,
n~non-relativistic. \\
Bond lengths $r$ and bond energies $\Delta E$
are in au.
\end{flushleft}
\end{ruledtabular}
\end{table}
\endgroup

Table~\ref{tab:mol} shows our tests
on a small but representative set of molecules
using the CCSD method and the L2\_4 basis set.
The bond lengths and bond energies
(without zero-point vibrations)
begin to differ between the nonrelativistic
and the four-component scalar-relativistic theories
on going to heavier atoms,
while our new two-component approximation
brings them close to the latter,
and although it becomes slightly worse for the heaviest atoms,
it is still rather good.


Our simple two-component formalism should be very well matched
to the density functional methods of generalized-gradient type~\cite{B88,PBE96}
and beyond~\cite{TPSS03,DRSLL04,VV10},
we have already implemented it
into our density-functional code~\cite{L97}
through analytic second derivatives.

\end{document}